# Supersonic Collisions between Two Gas Streams[1]

Hyung Mok Lee and Hyesung Kang
Dept. of Earth Sciences, Pusan National University, Pusan 609-735, Korea
email: hmlee@astrophys.es.pusan.ac.kr and kang@astrophys.es.pusan.ac.kr

Dongsu Ryu
Dept. of Astronomy & Space Sci., Chungnam National University, Daejeon 305-764, Korea
email: ryu@sirius.chungnam.ac.kr

## ABSTRACT

A star around a massive black hole can be disrupted tidally by the gravity of the black hole. Then, its debris may form a precessing stream which may even collide with itself. In order to understand the dynamical effects of the stream-stream collision on the eventual accretion of the stellar debris onto the black hole, we have studied how gas flow behaves when the outgoing stream collides supersonically with the incoming stream. We have investigated the problem analytically with one-dimensional plane-parallel streams and numerically with more realistic three-dimensional streams. A shock formed around the contact surface converts the bulk of the orbital streaming kinetic energy into thermal energy. In three-dimensional simulations, the accumulated hot post-shock gas then expands adiabatically and drives another shock into the low density ambient region. Through this expansion, thermal energy is converted back to the kinetic energy associated with the expanding motion. Thus, in the end, only a small fraction of the orbital kinetic energy is actually converted to the thermal energy, while most of it is transferred to the kinetic energy of the expanding gas. Nevertheless the collision is effective in circularizing the debris orbit, because the shock efficiently transforms the ordered motion of the streams into the expanding motion in directions perpendicular to the streams. The circularization efficiency decreases, if two colliding streams have a large ratio of cross sections and a large density contrast. But even in such cases, the main shock extends beyond the overlapping contact surface and the high pressure region behind the shock keeps the stream of the larger cross section from passing freely. Thus the stream-stream collisions are still expected to circularize the stellar debris rather efficiently, unless the ratio of the cross sections is very large (i.e., $\sigma_1/\sigma_2 \gg 10$).

*Subject headings:* hydrodynamics – shock waves – galaxies : nuclei

---

## 1. Introduction

It is believed that the tidal disruption of stars can occur at a rate of approximately one in $10^4$ years in typical nuclei of nearby galaxies with observational evidence for a super massive black hole (*e.g.*, Rees 1988; Cannizzo, Lee & Goodman 1990). A large amount of energy ($\sim 0.1 \, m_* c^2$) can be released if a significant fraction of stellar mass is accreted onto the black hole. However, galaxies with convincing dynamical evidence for a central black hole usually do not reveal strong activities in their central parts (*i.e.*, large $M/L$). It is possible that the dynamical mass implied by spectroscopic observations might not be due to the massive black hole but a cluster of dark stars (Goodman & Lee 1989). High resolution observations in the near future would enable us to eliminate some candidates for the constituents of dark clusters. It is, however, too early to rule out the existence of a central massive black hole on the grounds of low-level activities in galactic nuclei. We simply might not understand well enough the physical processes of the tidal disruption of stars and the subsequent evolution of the debris.

One of the key questions here is how the stellar debris forms an accretion disk around the black hole and is accreted onto it (Rees 1994). Since the initial orbit of the debris is extremely eccentric, the accretion time scale is longer than the typical interval between two successive tidal disruption events (Gurzadyan & Ozernoy 1980, 1981). But, if the orbit of the debris can be circularized, the accretion time scale can become as short as a few tens of years (Cannizzo, Lee & Goodman 1990). It has been shown that strong relativistic effects make the orbit of the debris stream precess at a large rate around the black hole (on the order of one radian per revolution: see, for example, Rees 1988; Cannizzo, Lee & Goodman 1990; Monaghan & Lee 1994; Kochanek 1994). As a result, the parts of the stream with different orbital periods (we will call them the outgoing and incoming streams, hereafter) may cross each other, and this stream-stream collision may dissipate the orbital energy via a shock leading to an effective circularization.

So the circularization process through the stream-stream collision is a crucial issue in estimating the observational consequences of the tidal disruption of a star. However, the supersonic collision of two gas streams has not been investigated in detail, in order to tell us how the gas flow evolves subsequently and how accretion follows. One can expect that the details of the stream-stream collision and the efficiency of the circularization will depend upon the geometry and flow parameters of the gas streams at the time of the collision. Kochanek (1994) estimated that the widths and heights of the two colliding streams would be different by a factor of two or more, and the collision angle between the streams is typically 130° to 140°. He suggested that most of the larger stream would freely move away along its original path, because of significantly different cross sections, so that the stream-stream collision should have only small effects on the orbital evolution.

Motivated by this, we have studied the evolution of the collision of the outgoing and incoming streams and estimated the degree of the conversion of the stream's orbital kinetic energy into other forms of energy. First, we have studied analytically the collisions of two one-dimensional, semi-infinite streams and showed how the shock parameters depend on the stream density and velocity.

In more realistic situations, the cross sections of the two colliding streams can be very different and also the collision is not necessarily head-on. Thus we have performed three-dimensional numerical simulations for the collisions of two supersonic streams with different density ratios, cross section ratios, and collision angles.

In §2 we present the one-dimensional analytic study. This is followed by the description of the three-dimensional numerical study in §3. In §4 we describe the results and implications of the simulations. Conclusions are given in §5.

## 2. One-dimensional Plane-Parallel Collision

In this section, we describe the evolution of the collision of two plane-parallel gas streams, for which we can find a simple solution. It tells us some essential features found in the three-dimensional simulations.

Suppose that two one-dimensional, semi-infinite, gas streams have density, $\rho_1$ and $\rho_2$, pressure, $p_1$ and $p_2$, and velocity, $v_1 > 0$ and $v_2 < 0$ and that they collide at $t = 0$. This is an example of the Riemann problem where a gas flow containing an arbitrary discontinuity sets into motion. Fig. 1 shows the solution of the case where $\rho_1 = 4$, $\rho_2 = 1$, $p_1 = p_2 = 2.4 \times 10^{-6}$, and $v_1 = -v_2 = 0.2$. The adiabatic index is $\gamma = 5/3$. The solution for the post-shock regions 3 and 4 can be found from the intersection of the two Hugoniot curves on the so-called $(p, v)$ diagrams for the two shocks (see, e.g., Zel'dovich & Raizer 1966). Here, the two shocks propagate from both sides of the contact discontinuity (CD). The pressure and velocity are continuous across the contact discontinuity, i.e., $p_3 = p_4$ and $v_3 = v_4 = v_{CD}$, where $v_{CD}$ is the velocity of the contact discontinuity.

Since the shocks are strong (i.e., $p_3$, $p_4 \gg p_1$, $p_2$, so $\rho_3/\rho_1 = \rho_4/\rho_2 = (\gamma+1)/(\gamma-1)$), we find the following simple relations:

$$p_3 = p_4 = \frac{\gamma + 1}{2} \frac{\rho_1 \rho_2}{(\sqrt{\rho_1} + \sqrt{\rho_2})^2}(v_1 - v_2)^2 \tag{1}$$

$$v_3 = v_4 = v_{CD} = \frac{\sqrt{\rho_1} v_1 + \sqrt{\rho_2} v_2}{\sqrt{\rho_1} + \sqrt{\rho_2}} \tag{2}$$

$$u_1 = -\frac{\gamma + 1}{2} \frac{\sqrt{\rho_2}}{\sqrt{\rho_1} + \sqrt{\rho_2}}(v_1 - v_2), \tag{3}$$

$$u_2 = \frac{\gamma + 1}{2} \frac{\sqrt{\rho_1}}{\sqrt{\rho_1} + \sqrt{\rho_2}}(v_1 - v_2). \tag{4}$$

Here $u_1$ and $u_2$ are the shock velocities relative to the upstream flows in the regions 1 and 2, respectively. The ratio of the shock velocities is $|u_1/u_2| = \sqrt{\rho_2/\rho_1}$, so the ram pressures by the two shocks are same (i.e., $\rho_1 u_1^2 = \rho_2 u_2^2$).

One can define the *thermal energy conversion flux*, $F_{th}$, as the postshock thermal energy generated by the shock per unit area per unit time according to

$$F_{th} = \frac{p_3}{(\gamma - 1)}(w_2 - w_1). \tag{5}$$

Here $w_1 = u_1 + v_1$ and $w_2 = u_2 + v_2$ are the shock velocities in the frame where the initial flows are defined. In the strong shock limit,

$$F_{th} = \frac{\gamma + 1}{4} \frac{\rho_1 \rho_2}{(\sqrt{\rho_1} + \sqrt{\rho_2})^2}(v_1 - v_2)^3 \tag{6}$$

Then, the *conversion ratio*, $R$, can be defined by the ratio of the thermal energy conversion flux to the influx of the kinetic energy; that is,

$$R = \frac{F_{th}}{(1/2)\rho_1 v_1^3 + (1/2)\rho_2 v_2^3}. \tag{7}$$

The probable configuration for the collisions of the debris streams mentioned in §1 is the one with $\chi = \rho_1/\rho_2 \gtrsim 1$ and $v_2 \sim -v_1$. In this case, $|u_1/u_2| = \sqrt{\chi}$. Then the postshock pressure is given by

$$p_3 = \frac{\gamma + 1}{2}\rho_2(v_1 - v_2)^2 \frac{\chi}{(\sqrt{\chi} + 1)^2}, \tag{8}$$

if the shocks are strong. The term $\chi/(\sqrt{\chi}+1)^2$ varies from $1/4$ to $1$ for $\chi \geq 1$. Thus, the postshock pressure is determined mostly by the relative velocity of the gas streams and the density of the lower density stream, but it is almost independent of the density of the higher density stream. Since $F_{th} \propto p_3(v_1 - v_2)$, the thermal energy conversion flux is also determined by the density of the lower density stream. The conversion rate $R$ decreases with increasing $\chi$, so the energy conversion is less efficient if the density contrast is large. In the limit of high $\chi$ ($\chi \to \infty$), the collision of two streams reduces effectively to a problem where a solid body moving into a low density stream. Then, only one *bow* shock propagates into the low density medium and the postshock pressure is the ram pressure on the solid body. The contact surface moves with the velocity of the solid body.

Another extreme is case the one where the colliding gas streams are identical ($\chi = 1$) and have opposite velocities, $v_2 = -v_1$. Then the postshock regions have $p_3 = p_4 = (4/3)\rho_1 v_1^2$ and $v_3 = v_4 = 0$, and the shock speeds relative to the upstream flows are $u_1 = -u_2 = -(4/3)v_1$. The thermal energy conversion flux is $F_{th} = (4/3)\rho_1 v_1^3$, and the conversion ratio is $R = 4/3$ which is the largest. The reason why $R$ is greater than one can be understood as follows. All the kinetic energy of the gas that passes through the shock is converted into the thermal energy, since the postshock gas is at rest in the lab frame. But the shocks move outwards from the contact point, so $R = |u_1/v_1|$ which is $4/3$ for $\gamma = 5/3$.

## 3. Three-dimensional Hydrodynamic Simulations

We are interested in the situation where the gas streams move into an extremely low-density and low-pressure medium with a supersonic speed. For numerical simulations, however, we have adopted the flow parameters that can be handled comfortably with our code. In the collisions of two identical streams, we have set $\rho = 1$, $p = 2.4 \times 10^{-6}$, and $v = 0.2$ for the streams, so the Mach number is $v/c_s = 100$. Since the colliding gas is cold, the collision with a higher Mach number would represent a more realistic situtation. We have chosen the Much number 100, since it is high enough but still can be handled comfortably with our code. It is expected that the streams of stellar debris may have cross sections inversely proportional to the density, $i.e.$, $\sigma \cdot \rho = $ constant, while they have similar orbital speeds (Kochanek 1994; Monaghan & Lee 1994). Thus in the collisions of two different streams, we vary the density and the cross section according to $\sigma \cdot \rho = $ constant. We use $\rho_2 = 1$ for the low density stream and the density contrast is $\chi = \rho_1/\rho_2 = \sigma_2/\sigma_1 = 4$, 9, or 16. The speed and pressure are the same, that is, $v_1 = v_2 = 0.2$ and $p_1 = p_2 = 2.4 \times 10^{-6}$. For the density contrasts considered here ($\chi =$4 to 16), we expect that the results would be somewhere between those of the identical streams ($\chi = 1$) and those of a high density contrast ($\chi \to \infty$).

The density and pressure of the ambient medium are set to be $\rho_b = 10^{-3}\rho_1$, and $p_b = 10^{-3}p_1$. Since the pressure of the stream is higher than that of the ambient medium, a shock propagates into the ambient medium and a rarefaction wave into the stream. But the shock is weak with the Mach number $u_{ps}/c_b = 1.8$, where $c_b$ is the sound speed of the ambient medium. If the ambient medium is a vacuum, the shock has the maximum speed $u_{ps} = 2c_b/(\gamma - 1)$. However, the maximum shock speed is still much smaller than the stream speed, so the shock propagating from the stream boundary is not dynamically important at all.

For the boundary condition of the streams, we have used the inflow condition in which the flow variables of the streams entering the computational domain are continuous across the boundary ($i.e.$, no gradient). Such inflow condition can be handled exactly, only when the inflow angle relative to the boundary is given by $\tan\theta_{\rm in} = n_1/n_2$ where $n_1$ and $n_2$ are integers from 0 to $\infty$. Hence, in our simulations we consider only collisions with collision angles given by $\tan(\theta/2) = 1/3$, $1/2$, 1, 2, 3, and $\infty$. The most likely collision angle between the outgoing and incoming gas streams expected around a black hole is greater than 90°; more restrictively between 130° and 140° as mentioned in §1. So we pay special attention for the case of $\theta = 143°$ and 180°.

We divide the simulations into two groups. Group A is for the simulations with two identical streams colliding with different collision angles. So Model "A36", for example, is for the collision of two identical streams with the collision angle 36°. Group B is for the simulations with two different streams colliding with either 143° or 180°. The stream parameters, such as the ratios of the densities and radii of the cross sectional areas of the two streams, are given in Table 1.

The simulations have been done with a three-dimensional hydrodynamics code based on the Total Variation Diminishing (TVD) scheme (Harten 1983; Ryu et al. 1993). The TVD scheme is an explicit, second-order, Eulerian finite difference scheme. In the simulations with the TVD code, shocks spread over 2 − 3 cells. In our simulations the computational box is a unit cube typically with $128^3$ cells. Since the fast stream has a speed 0.2, it can cross the box in $t = 5$. The streams has been set to collide at $t = 0$ and the temporal integration have been done up to $t = 2.0$. A typical run takes less than $\sim 1000$ CPU seconds on a Cray C90.

## 4. Results and Implications

### 4.1. Morphology

Let us start by overviewing the morphological shape in the stream-stream collisions. To see some of the key features described in the following, refer to Figs. 2 to 5. Fig. 2 shows the pressure contour maps of a slice cut through the middle plane of the colliding streams for four models in Group A (A18, A36, A90, and A180). Fig. 3 shows the pressure contour maps for models of $\theta = 143$ (*i.e.*, A143, B143a, B143b and B143c). The density contour maps for the same models as shown in Fig. 3 are plotted in Fig. 4. Fig. 5 shows the velocity maps for models A143, B143c, A180, and B180. Morphologically the most noticeable feature is the oblique shocks formed around the contact surface. They can be identified by strong gradients in the pressure contour maps.

For the collisions of two identical streams, one can define a symmetry plane which contains the initial contact surface. The angle between the oblique shocks and the symmetry plane depends upon the flow parameters and the collision angle. This angle is zero for the collision angle $\theta = 180°$ and increases up to a maxim angle $\sim 11°$ as $\theta$ decreases to 90°. Then it decreases, as the collision angle further decreases below 90°. Once the angle between the oblique shocks and the symmetry plane is known, the shock jump condition can be found by applying the usual Hugoniot-Rankine relations. As the collision angle increases, the velocity component perpendicular to the symmetry plane increases so that the postshock gas pressure behind the oblique shock increases.

The collision of two identical streams with the collision angle $\theta = 180°$ (A180) is compared most closely to the one-dimensional collisions considered in the previous section. In the one-dimensional collisions, the shocks propagate back to the upstream flows and the shock speed relative to the upstream flows is $4/3v_{st}$. But in the three-dimensional collisions the shock speed is $\sim v_{st}$, since the shocks stand off near the symmetry plane instead of penetrating into the upstream flows. This is because the high pressure behind the shocks pushes the shock gas into the low pressure ambient medium in the direction perpendicular to the stream flows. The shocked gas is first accumulated into a dense region between the shocks, and then accelerated into the ambient medium. This expanding hot gas drives another shock in the ambient medium. This secondary shock is identified by strong pressure gradients in the background medium. The speed of this shock is similar to the stream speed itself, since the postshock pressure is similar to the ram pressure of the streams. The high pressure shocked region is roughly divided into a small volume of the densest gas surrounded by main shocks and a large volume of diffuse expanding flow encompassed by the secondary shock.

In the cases with the collision angle greater than 90° but less than 180° (*e.g.*, A143), a forward shock and a reverse shock propagate into the two directions in the ambient medium. The forward shock moves faster than the reverse shock due to the positive bulk motion of the streams along the symmetry plane. The three-dimensional shape of the expanding high pressure region is then a torus that encompasses the contact surface, but the cross section of the torus is not symmetric. When the collisional angle is smaller than 90° (see Fig. 2), the reverse shock is absent.

In the collisions of two different streams (B143abc and B180), the high density stream penetrates somewhat into the low density stream since the high density stream has a smaller cross section. However, the oblique shocks extend beyond the overlapping contact surface which is determined by the cross section of the high density stream. The flow outside the overlapping contact surface is prevented from streaming freely by the extended oblique shocks and the torus of the high pressure gas. Hence, contrary to a naive expectation, the geometric cross section of interacting streams seems to have a rather weak influence on the dynamics of the stream-stream collisions.

### 4.2. Energy Conversion

The component of the orbital kinetic energy of the streams due to the motion perpendicular to the oblique shocks is converted to the thermal energy by heating the shocked gas. But the heated gas expands into the ambient medium and cools adiabatically. Through this adiabatic expansion, the thermal energy is converted back to the kinetic energy of the expanding gas encompassed by the secondary shocks, as described in the previous subsection. In order to quantify the efficiency of the conversion of the stream's orbital kinetic energy into the thermal energy, we define the energy conversion rate, $R$, as the ratio of the increase in the total thermal energy in the computational box to the kinetic energy added to the box through the boundary, *i.e.*,

$$R = \frac{U_{tot}(t) - U_{tot}(0)}{(1/2)\rho_1 v_1^2(\sigma_1 v_1 t) + (1/2)\rho_2 v_2^2(\sigma_2 v_2 t)}, \qquad (9)$$

where $U_{tot}(t)$ is the total thermal energy at $t$. This ratio has a similar meaning to the conversion rate defined in Eq. (7). $R$'s for various models are plotted against $t$ in Fig. 6. The slow increase in $R$ during the early phase for $\theta$ less than 180° is due to the fact that the area of the contacting surface increases with time. The subsequent decrease is due to the following reason. The total thermal energy $U_{tot}(t)$ increases initially with $t$, but then stays more or less constant after the contact surface is fully developed. But the kinetic energy in the denominator of Eq. (9) continues to increases with time, so the ratio $R$ decreases at late epochs. In the end of our simulations ($t = 2$), only a small fraction of the stream's orbital kinetic energy ($\lesssim 30\%$) was converted to thermal energy and most of it went to the expanding kinetic energy of the hot torus gas.

Since the initial orbital energy of the streams is converted eventually to the expanding energy of the shocked gas, the nature of the debris orbit would change drastically after the stream crossing. The amount of energy transferred to the expanding motion can be estimated from the maximum value of the energy conversion rate. If the maximum $R$ is close to 1, almost all the stream's orbital energy will be eventually converted into the expanding energy. In the cases of the collisions of two identical streams, the maximum $R$ is generally close to 1. The maximum $R$ becomes smaller as the cross sectional area (and thus density) contrast grows, but the dependence is rather weak. For example, the maximum $R$'s of the different stream cases (for $\chi = 4 - 16$) are about $30 - 35\%$ of that of the identical stream case (see Fig. 6c). Thus the conversion of the orbital energy to the expanding energy is still significant unless the density contrast is very large (i.e., $\chi \gg 10$). This is mainly due to the fact that the oblique shocks affect the area beyond the overlapping contact surface, as discussed in the previous subsection. Similar level of decrease in the energy conversion with the density contrast is seen from Fig. 6d ($\theta = 180°$) where head-on collision cases are shown.

### 4.3. Implications on the Long Term Evolution of Stellar Debris

The simulations of colliding two streams would comprise of only a small part of the evolution of stellar debris around a black hole, but we may still extract some important physics. In this subsection, we attempt to predict how the collision of debris streams affects their evolution and how the circularization of the debris orbit proceeds, by relating the results of the present simulations to those of stellar disruption simulations described in Monaghan & Lee (1994).

In Fig. 7, we show the typical geometry of gas streams resulted from an SPH simulation for the disruption of a star near a black hole (Monaghan & Lee 1994). The arrows indicate the direction and the size of the velocity of SPH particles. The magnitude of the velocity of the streams is close to escape velocity, because the eccentricity is almost 1. Therefore, at a distance $r$ from the black hole it can be approximated as

$$v \approx \left(\frac{2GM}{r}\right)^{1/2}, \qquad (10)$$

where $M$ is the mass of the black hole. The angle of the stream crossing is approximately given as

$$\theta \approx \pi \left(1 - 1.5 \frac{R_S}{r_p}\right), \qquad (11)$$

where $r_p$ is the pericentral distance of the orbit of the disrupted star and $R_S$ is the Schwarzschild radius of the black hole. Here we have assumed that the velocity vector at the crossing is nearly *radial*. The relation given by Eq. (10) can be confirmed from Fig. 8, where we have shown the velocity of SPH particles as a function of distance from the black hole. Note that the length unit shown in Fig. 7 and 8 is $R_S/2$. The circles denote the outgoing particles while the crosses represent the incoming particles. There is no distinction in orbital velocities for the incoming and outgoing particles. Since the location of stream crossing is about $r_{cr} = 50R_S$ (or 100 in present length unit) from the black hole, it can be seen that the velocities are closely approximated by Eq. (10). With a typical value of $r_p \approx r_{tid} \approx 5R_S$, we obtain $\theta \approx 130°$, where $r_{tid}$ is the tidal disruption distance.

We can estimate the average velocity of the shocked streams along the tangential direction as $v_{tan} \sim v_{st}\sqrt{2\cos(\theta/2)}$ for the collision of two identical streams, if we assume that the collision is *sticky* (*i.e.*, inelastic) and the gas is incompressible. It is equal to the circular velocity of $v_{cir} = \sqrt{GM/r}$ for $\theta \approx 151°$. Thus for the typical angle of $\theta \approx 130°$, the tangential velocity of the sticky, incompressible fluid will be close to the value for a circular orbit. In such case the orbital energy would be removed effectively and the shocked gas would move nearly circular orbit at the orbital radius $r_{cr}$. In reality, however, the gas is not sticky nor incompressible, so the shocked gas goes through an adiabatic expansion.

The geometry of the crossing debris streams would resemble that somewhere between A143 and B143c (see Figs. 3 and 4). In Fig. 9 we show the accumulated frequency distribution of the kinetic energy per unit mass ($E_k$) for all the gas within our simulation box at the end of our runs of models A143a and B143c. The kinetic energy of the initial streams ($E_{k,s}$) is indicated as a vertical line at $E_k/E_{k,s} = 1$. The presence of gas at both lower and higher sides of the initial $E_k$ is due to the shock. Bulk of the shocked gas have energy smaller than the initial stream's kinetic energy. Thus the stream crossing will make the shocked gas more tightly bound to the black hole. A small fraction of the shocked gas has kinetic energy greater than that of the initial streams, as evident from the high velocity tail in Fig. 9. The amount of gas having $E_k > E_{k,s}$ is about 1/4 of that with $E_k < E_{k,s}$. However, still majority of the high velocity gas lies very close to $E_{k_s}$, so the fraction of gas that will be unbound from the black hole is likely to be very small (i.e., < 10%).

The difference in the circularization efficiency between the identical and different stream collisions can be noticed also from Fig. 9. About half of the gas with $E_k < E_{k,s}$ have $E_k < 0.5E_{k,s}$ for the model A143a, while the fraction becomes less than a third for the model B143b. $E_k = 0.5E_{k,s}$ represents approximately the energy of the gas having a circular orbit after the shock. Thus the particles with $E_k < 0.5E_{k,s}$ have a apocenter smaller than $r_{cr}$, while those with $E > 0.5E_{k,s}$ have a apocenter greater than $r_{cr}$.

Once the stream crossing begins, the incoming stream no more exists beyond the crossing point. The tail of the incoming stream moves around the black hole and becomes the endpoint of the outgoing stream. After the tail of the outgoing stream is destroyed by the collisions, the incoming stream will flow freely again. Thus the duration of a stream-stream collision until the exhaustion of one segment of the outgoing stream is about the time required for a particle to have a round trip from the location of the stream crossing through the apocenter from the black hole. For a particle with a nearly parabolic orbit, the duration is twice the free fall time. Thus the collision time scale would be $t_{col} \approx 2\sqrt{(3\pi^2 r^3)/(8GM)} \sim 50$ hours for our choice of typical parameters (i.e., $M = 10^7 M_\odot$, $r_{cr} \approx 50 R_S$, and $r_p \approx r_{tid} \approx 1$ AU). The quiescent phase would also have the similar duration. Therefore, the stream collision could be a periodic phenomena with period about 100 hours. A tidal disruption with smaller $r_p$ would produce shorter period, because $r_{cr}$ is smaller. Obviously, such a periodicity is less pronounced if the contrast in cross sectional areas becomes larger.

## 5. Conclusion

In this paper, we have investigated the supersonic collisions between two gas streams with various configurations and collision geometry. In the collisions of one-dimensional plane-parallel streams, the kinetic energy is converted effectively into thermal energy via the shocks which propagate into the streams. The conversion is most effective for the collisions of two identical streams. It becomes least effective if the two streams have an infinite density contrast. For the intermediate density contrast, the thermal energy conversion is proportional to the ram pressure of the low density stream but nearly independent of that of the high density stream. If the shocked gas can cool radiatively in a time scale shorter than the dynamical time, the kinetic energy of the streams would be dissipated more effectively.

The collisions of three-dimensional streams are much more complicated. We have more degrees of freedom in setting flow parameters such as the ratio of cross sections and collision angle. The key features of the three-dimensional simulations which contrast with those of the one-dimensional collisions include: 1) The shocks generated by the collisions stand near the stand-off point instead of penetrating into the colliding streams. 3) Most of the bulk kinetic energy of the streams turns into the kinetic energy of the expanding hot gas instead of the thermal energy. 2) The hot gas behind the shock expands uniformly in the directions perpendicular to the colliding streams.

In the collision of two identical streams, the fraction of the orbital kinetic energy converted to the expanding kinetic energy is high, so the circularization is effective. But realistic streams at the crossing point are likely to have a density contrast as large as $\sim 4 - 10$ and the ratio of cross sections is expected to vary as the reciprocal of the density contrast. In such cases, the conversion of the orbital kinetic energy should be somewhat smaller. If we quantify the circularization efficiency by the fraction of the orbital kinetic energy transferred to the expanding kinetic energy, the efficiency would be nearly 50% for B143a ($\chi = 4$) and B143b ($\chi = 9$) and 30% for B143c ($\chi = 16$), while nearly 100% for A143 ($\chi = 1$). The reason that the circularization efficiency depends only weakly on the ratio of cross sections is that the shock heated gas forms a *thick torus* of high pressure gas which keeps the stream of larger cross section from streaming freely. Thus stream-stream collision could still be an effective circularization mechanism, even if the ratios of densities and cross sections of two streams are significant ($\chi \lesssim 10$).

The detailed evolution of the expanding gas under the gravity of the black hole is still complicated. Some would escape from the black hole gravity, taking energy away with it and leaving the debris more tightly bounded to the black hole. Therefore, the collision could be considered as an effective way to dissipate the orbital kinetic energy of the streams, even in the absence of the radiative cooling of the shocked gas.

The stream crossing would last for a while (about 50 hours for typical parameters, i.e., $M = 10^7 M_\odot$, $r_{cr} \approx 50 R_S$, and $r_p \approx r_{tid} \approx 1$ AU) with a similar period of quiescent phase followed. Such a pattern is likely to repeat until the majority of stellar debris returns to the apocenter. Thus a periodic occurrence of *flare* could be observed if the stream crossing can produce any observable

– 11 –

phenomena.

So far, we have considered an adiabatic gas with the adiabatic index $\gamma = 5/3$ and ignored any radiative processes. However, in the collision of two gas streams, their kinetic energy could go mostly into radiation pressure rather than gas pressure depending on the parameters. Then $\gamma = 4/3$ would be more appropriate than $\gamma = 5/3$. But we believe that this is unlikely to make much of a difference to the outcome of our calculations. Also, it is possible that radiative processes might be important. For instance, if the torus of the shocked gas is optically thin, radiative losses could be efficient and remove internal energy almost as fast as it is created in the shocks. If it is optically thick, then radiative viscosity might play an important role. But since we have limited ourselves to the dynamics of an adiabatic gas in this paper, we leave these to future work.

We thank Drs. T. W. Jones and J. Goodman for comments on the manuscript. The work by HML was supported in part by KOSEF grant No. 941-0200-00102 and in part by the Cray R&D Grant in 1994. The work by HK was supported in part by the Korea Research Foundation through the Brain Pool Program. The work by DR was supported in part by the Basic Science Research Institute Program, Korean Ministry of Education 1995, Project No. BSRI-95-5408.

Table 1. Model Parameters

| Model | $\theta$ | $\rho_1/\rho_2$ | $r_1/r_2$ |
|:-:|:-:|:-:|:-:|
| A18 | 18 | 1 | 7/7 |
| A26 | 26 | 1 | 7/7 |
| A36 | 36 | 1 | 7/7 |
| A45 | 45 | 1 | 7/7 |
| A52 | 52 | 1 | 7/7 |
| A63 | 63 | 1 | 7/7 |
| A71 | 71 | 1 | 7/7 |
| A90 | 90 | 1 | 7/7 |
| A126 | 126 | 1 | 7/7 |
| A143 | 143 | 1 | 7/7 |
| A180 | 180 | 1 | 7/7 |
| B143a | 143 | 4 | 5/10 |
| B143b | 143 | 9 | 3/9 |
| B143c | 143 | 16 | 3/12 |
| B180 | 180 | 16 | 3/12 |

## REFERENCES

Cannizzo, J., Lee, H. M. & Goodman, J., 1990, ApJ, 351, 38.

Goodman, J., & Lee, H. M., 1989, ApJ, 337, 84.

Gurzadyan, V. G. & Ozernoy, L. M., 1980, A&A, 86, 315.

Gurzadyan, V. G. & Ozernoy, L. M., 1981, A&A, 95, 39.

Harten, A., 1983, J. Comp. Phys., 49, 357.

Kochanek, C. ApJ, 422, 508.

Monaghan, J. J. & Lee, H. M., 1994, in *Nuclei of Normal Galaxies*, NATO ASI Series C:445, eds. R. Genzel & A. I. Harris, (Kluwer), p449.

Rees, M. J., 1988, Nature, 333, 523.

Rees, M. J., 1994, in *Nuclei of Normal Galaxies*, NATO ASI Series C:445, eds. R. Genzel & A. I. Harris, (Kluwer), p453.

Ryu, D., Ostriker, J. P., Kang H., & Cen, R. 1993, ApJ, 414, 1

Zeldovich, Ya. B. & Raizer, Yu. P., 1966, *Physics of Shock Waves and High Temperature Hydrodynamic Phenomena*, Academic Press.

---

Fig. 1.— Flow structure after the collision of one-dimensional plane-parallel gas streams. The flow parameters $\rho_1, p_1, v_1$, and $\rho_2, p_2, v_2$ represent the initial condition left and right to the discontinuity. Two shocks propagate away from the discontinuity.

Fig. 2.— Pressure contour maps of a horizontal cut in models A18 (top left), A36 (top right), A90 (bottom left), and A180 (bottom right).

Fig. 3.— Pressure contour maps of a horizontal cut in models A143 (top left), B143a (top right), B143b (bottom left), and B143c (bottom right).

Fig. 4.— Density contour maps of a horizontal cut in models A143 (top left), B143a (top right), B143b (bottom left), and B143c (bottom right).

Fig. 5.— Velocity field maps of a horizontal cut in models A143 (top left), B143c (top right), A180 (bottom left), and B180 (bottom right).

Fig. 6.— Energy conversion rate $R$. Top two panels are for the models with two identical streams. $R$ increases with the collision angle. Bottom left panel is for the models with the collision angle 143°. The solid line is for A143 (i.e., identical streams), the dotted line for B143a ($\chi = 4$) the short dashed line for B143b ($\chi = 9$), and the dot-dashed line for B143c ($\chi = 16$). Bottom right panel is for the models with the collision angle 180°. The solid line is for A180 (i.e., identical streams), the short dashed line for B180 ($\chi = 16$).

Fig. 7.— A typical geometry of stream crossing resulting from an SPH simulation. The arrows indicate the locations and velocities of SPH particles. The length unit in this figure is $R_S/2$.

Fig. 8.— The velocity of SPH particles as a function of distance from the black hole for the case shown in Fig. 7. The straight line indicates the $v - r$ relationship for a parabolic orbit.

Fig. 9.— Accumulated frequency distribution of the kinetic energy per unit mass ($E_k$) for the entire gas in our simulation box at the end of simulations for models A143 and B143c. The kinetic energy of the incoming stream is indicated as a vertical line at $E_k/E_{k,s}=1$. Majority of shocked gas has lost its kinetic energy while a small fraction has gained it.

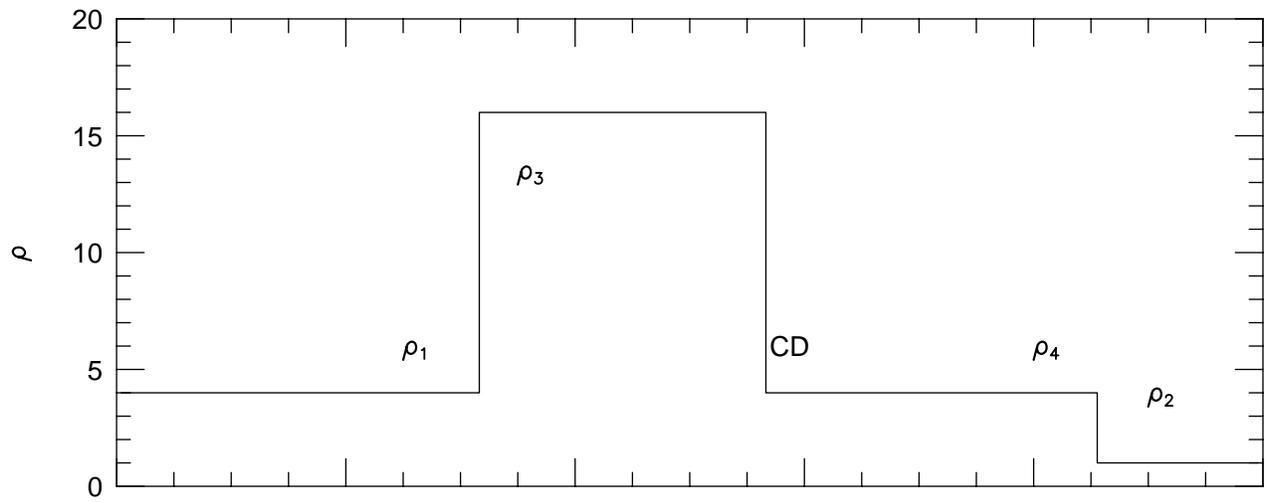

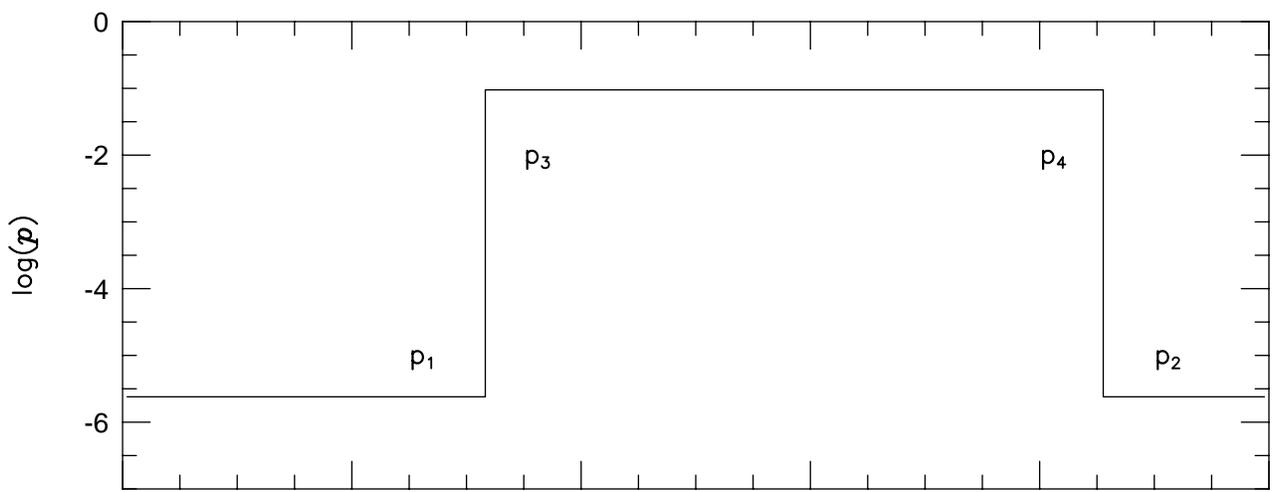

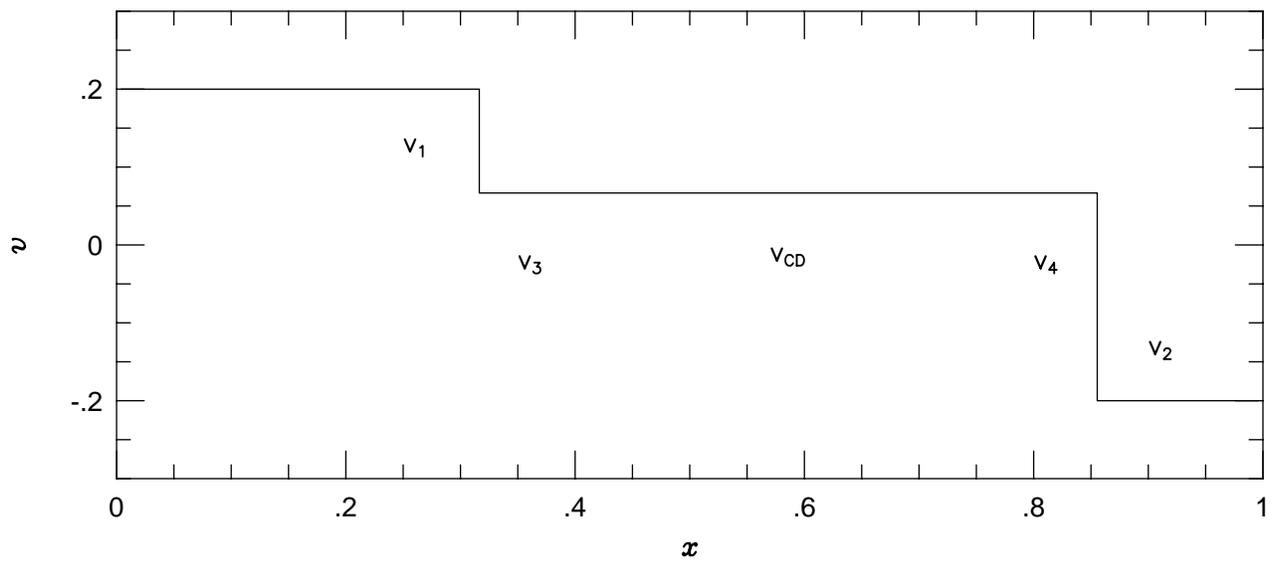

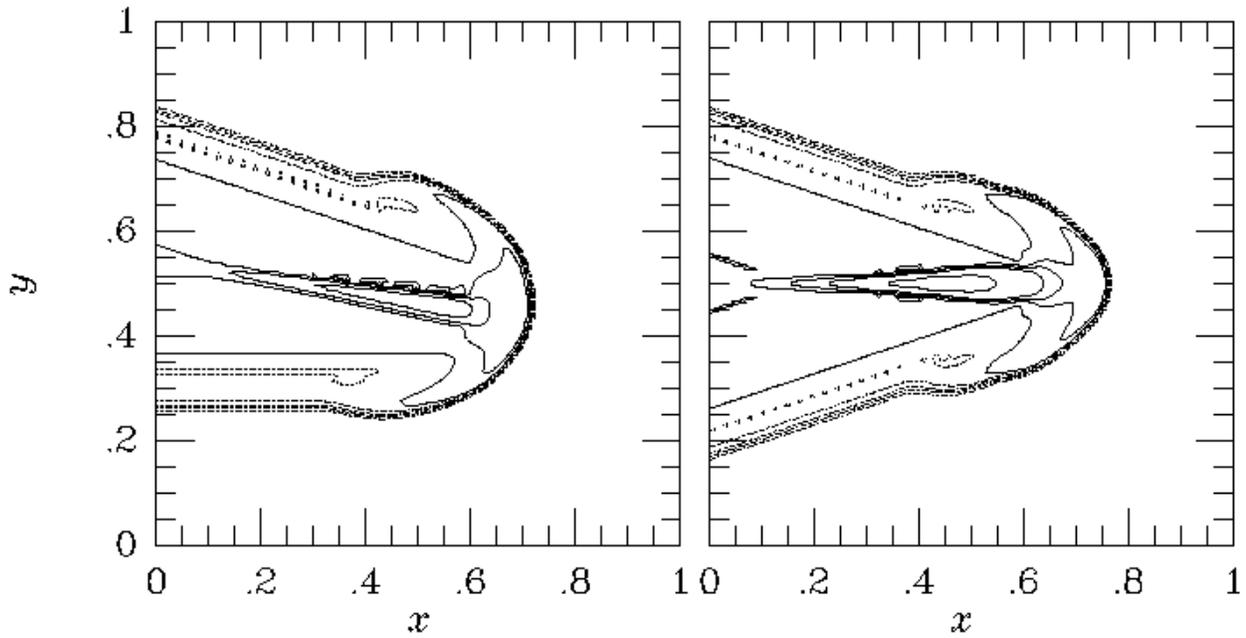
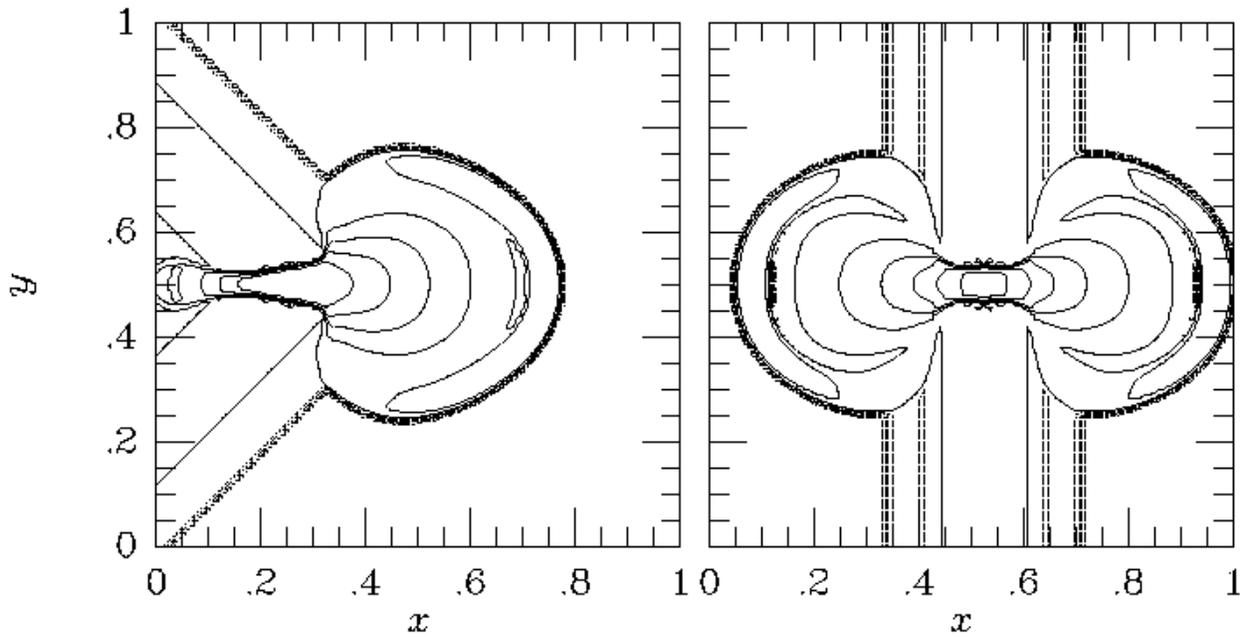

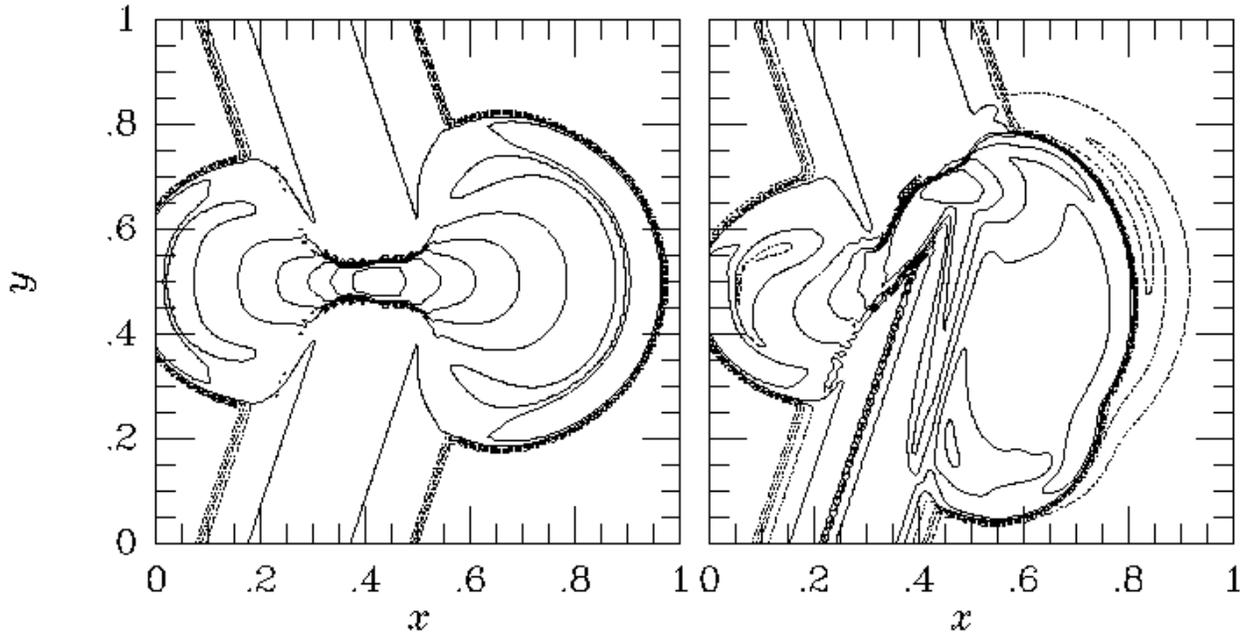
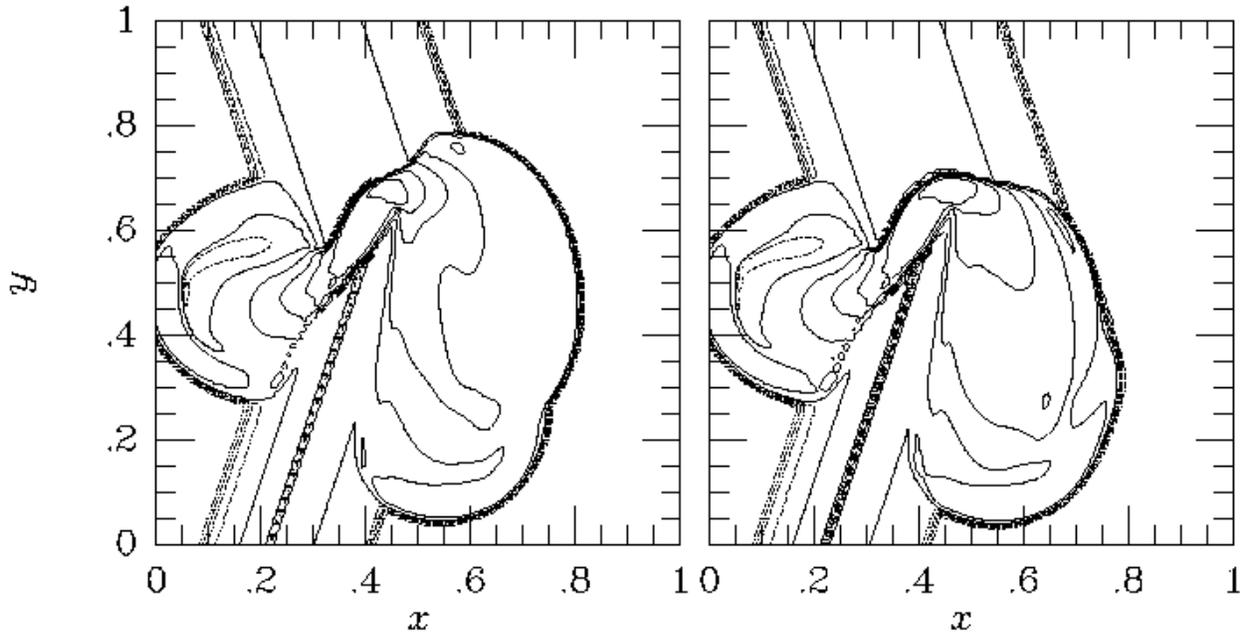

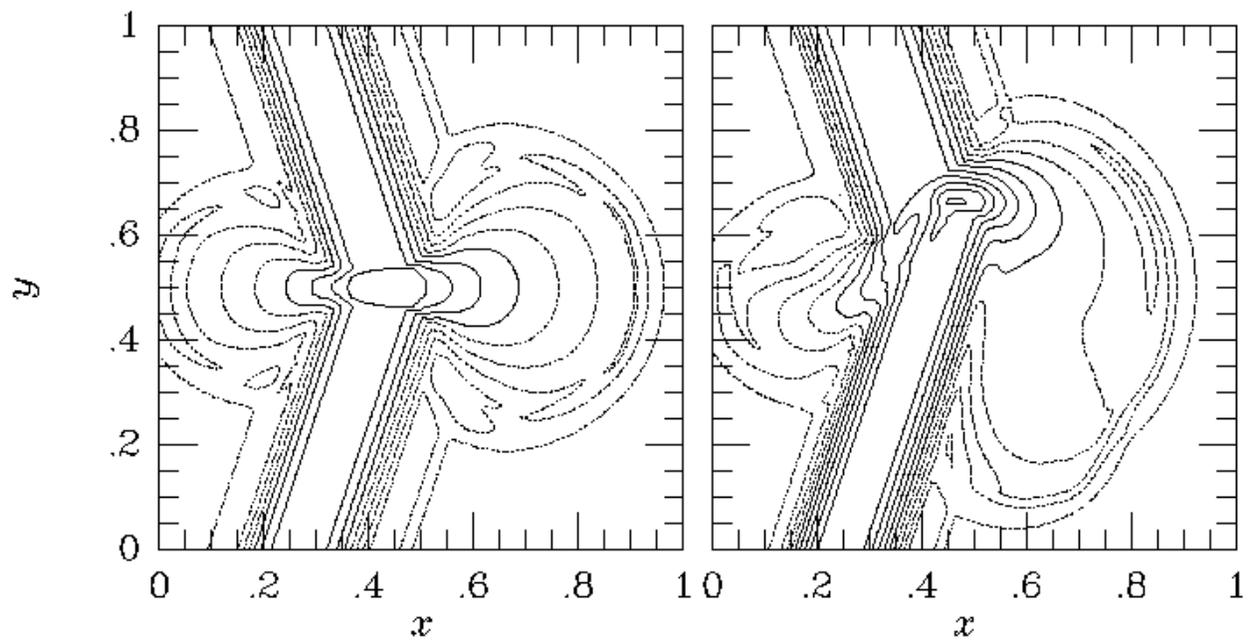
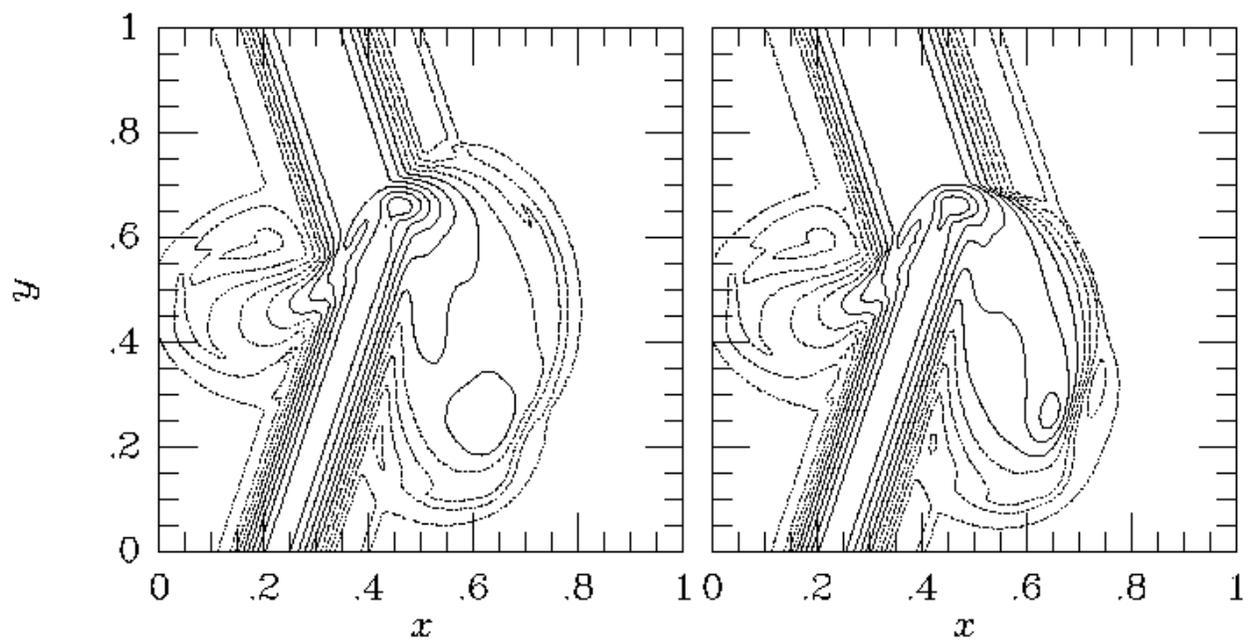

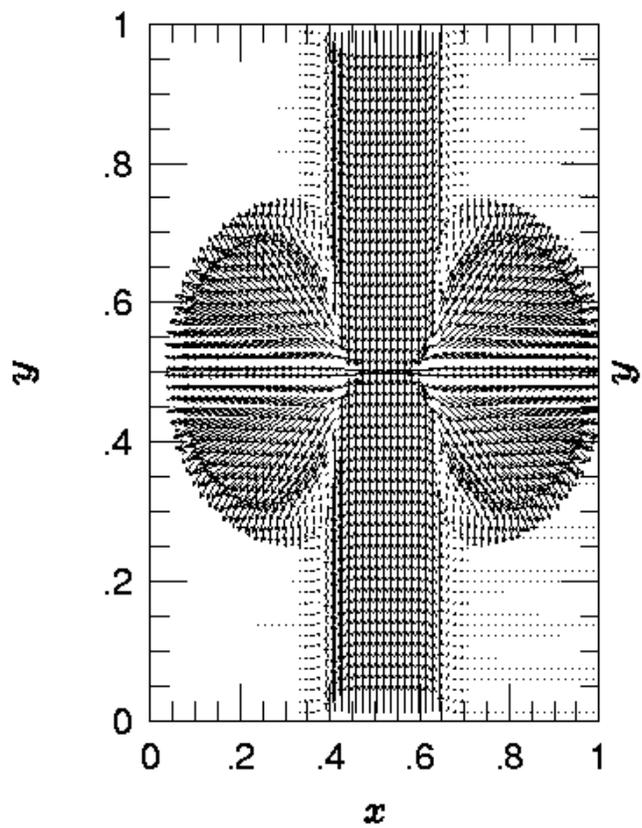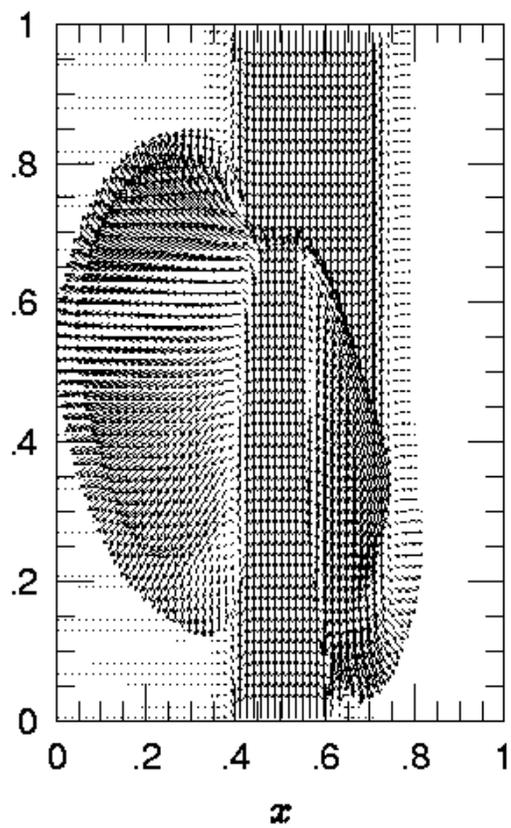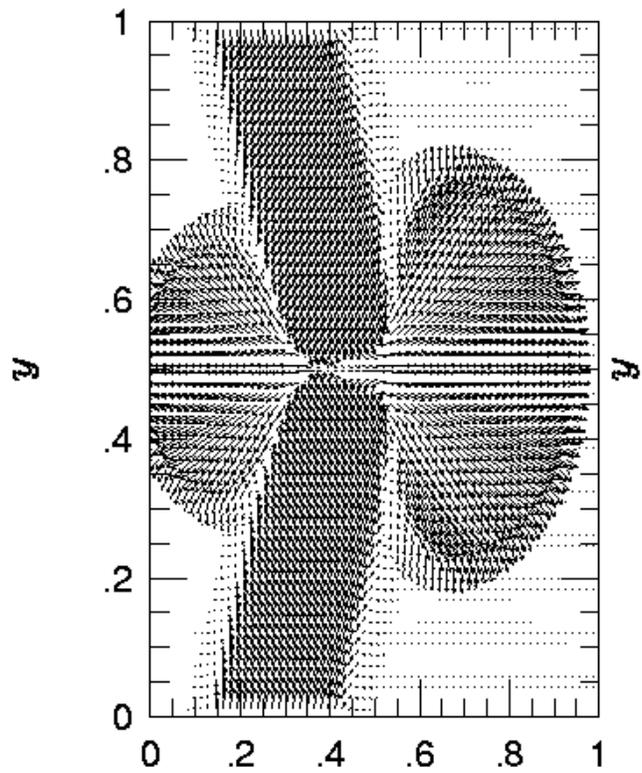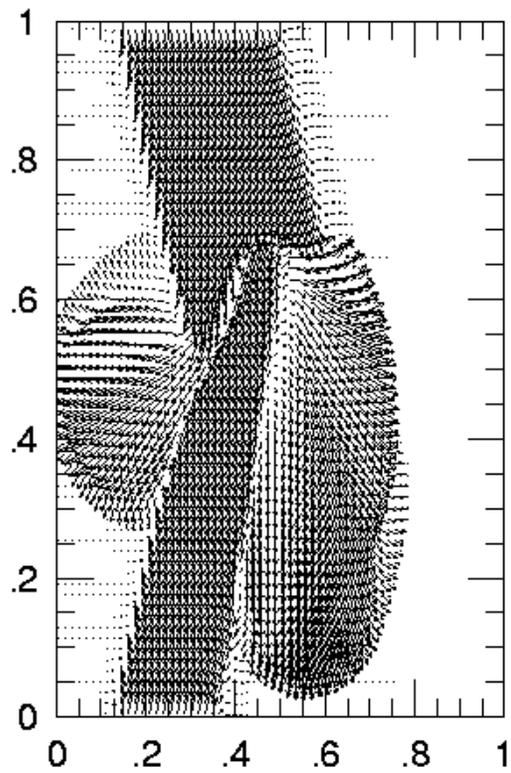

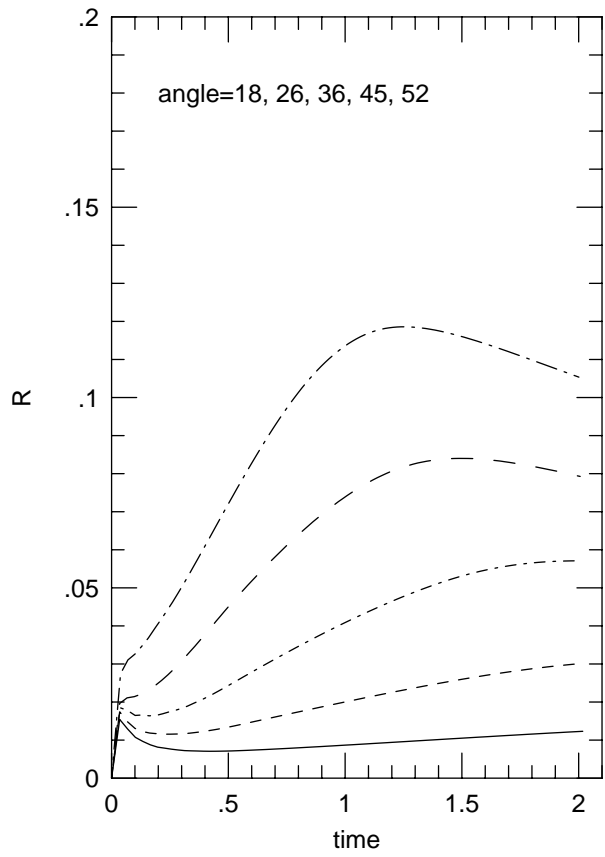
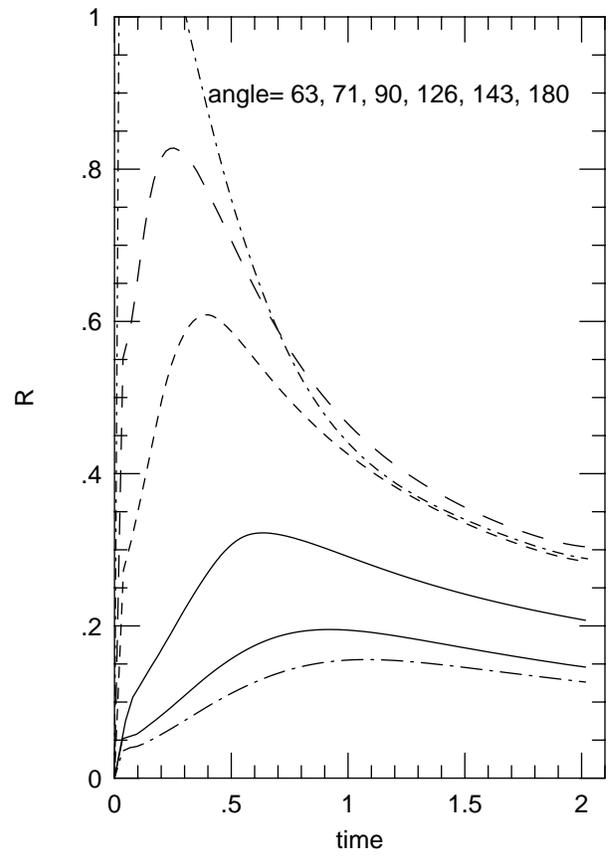
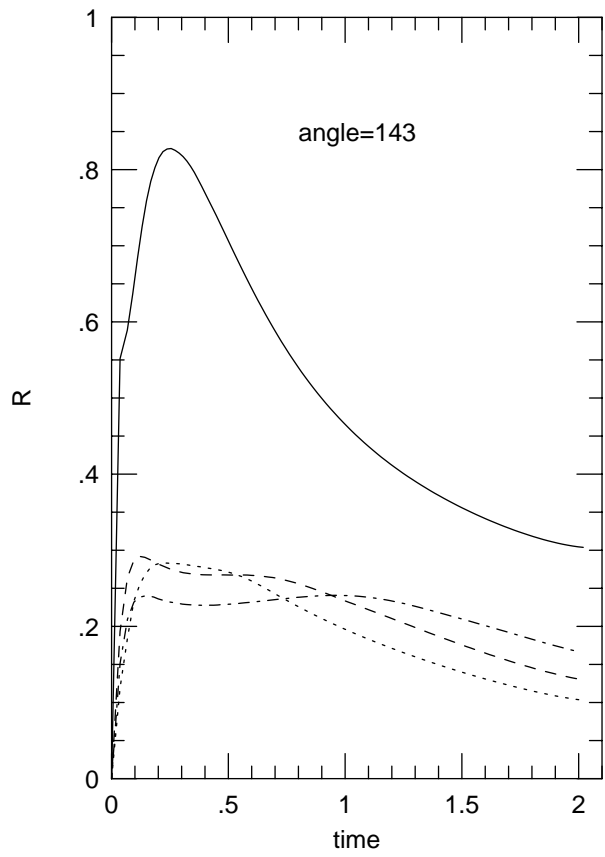
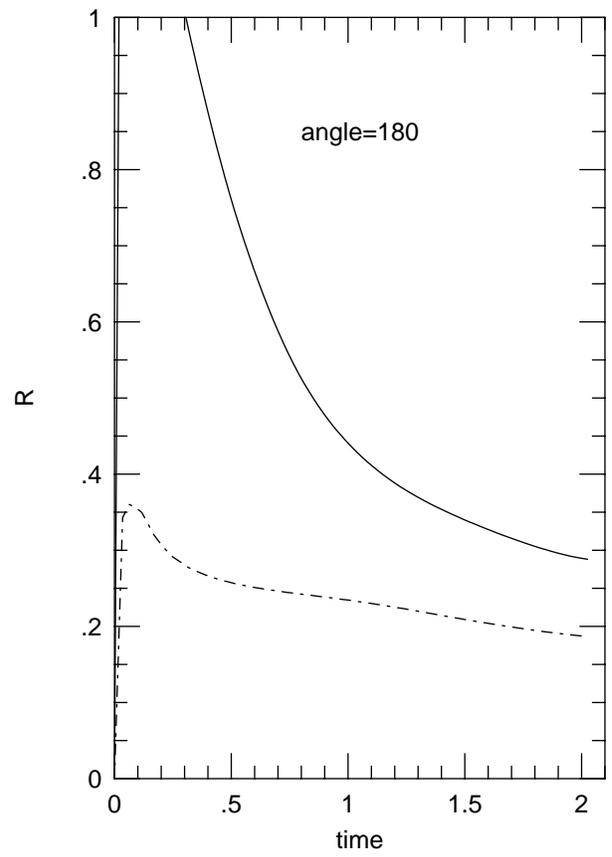

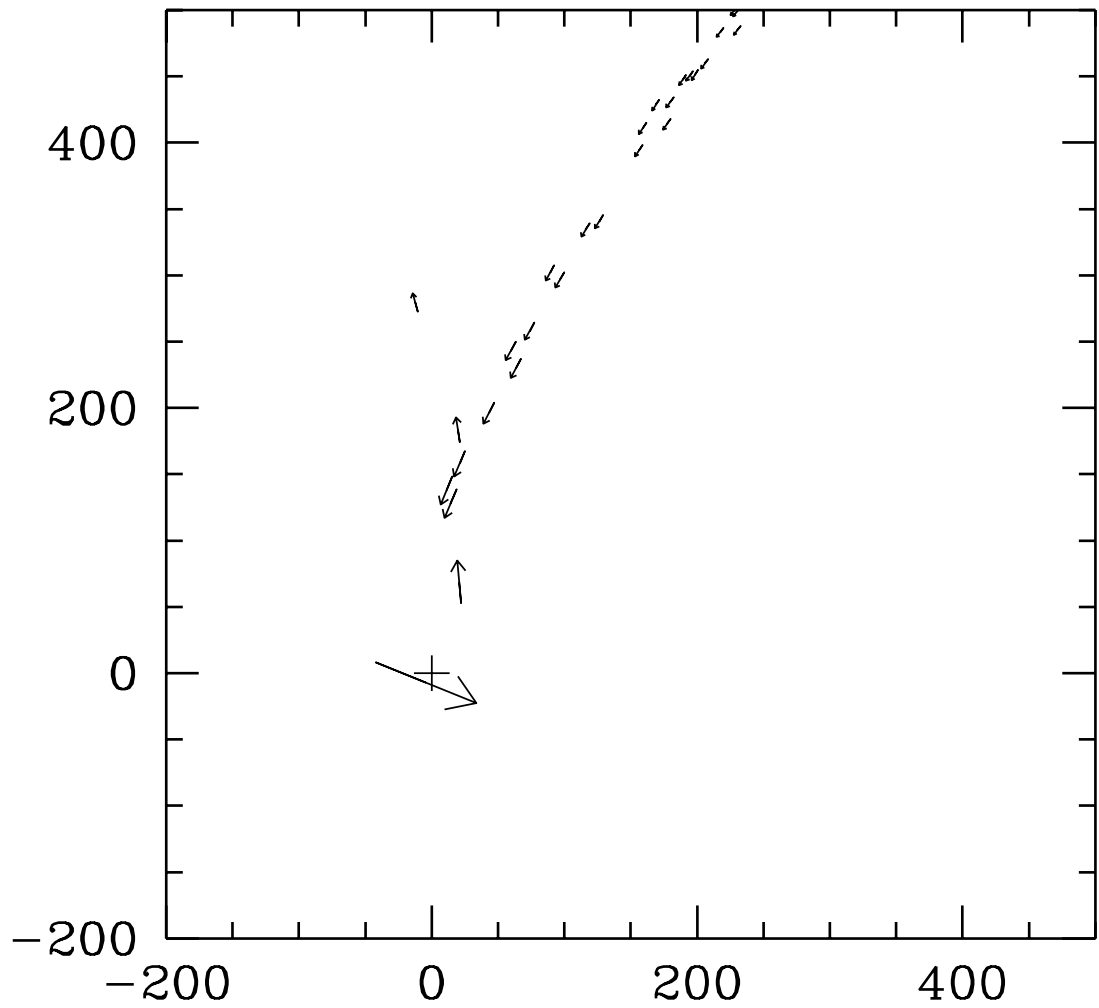

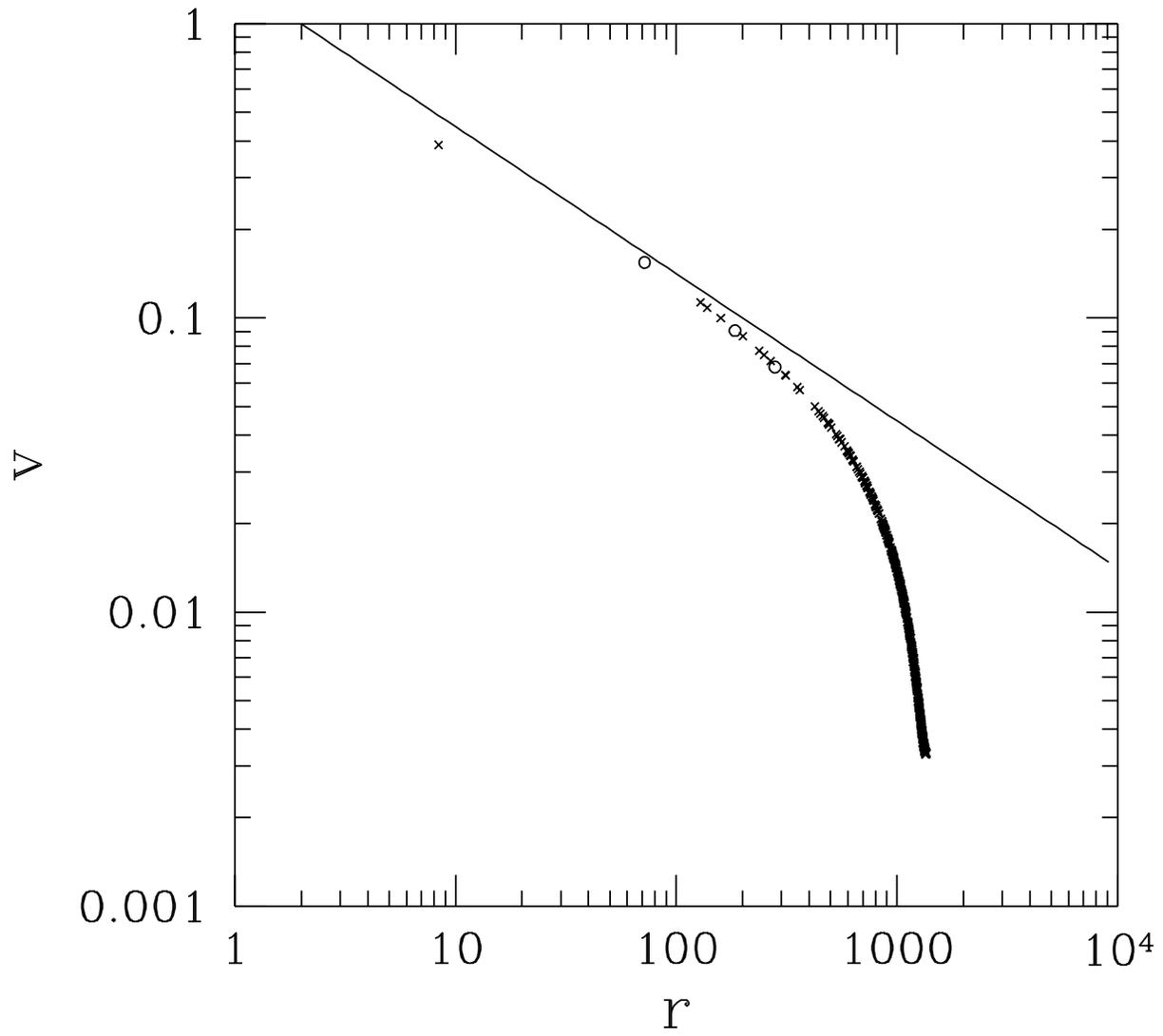

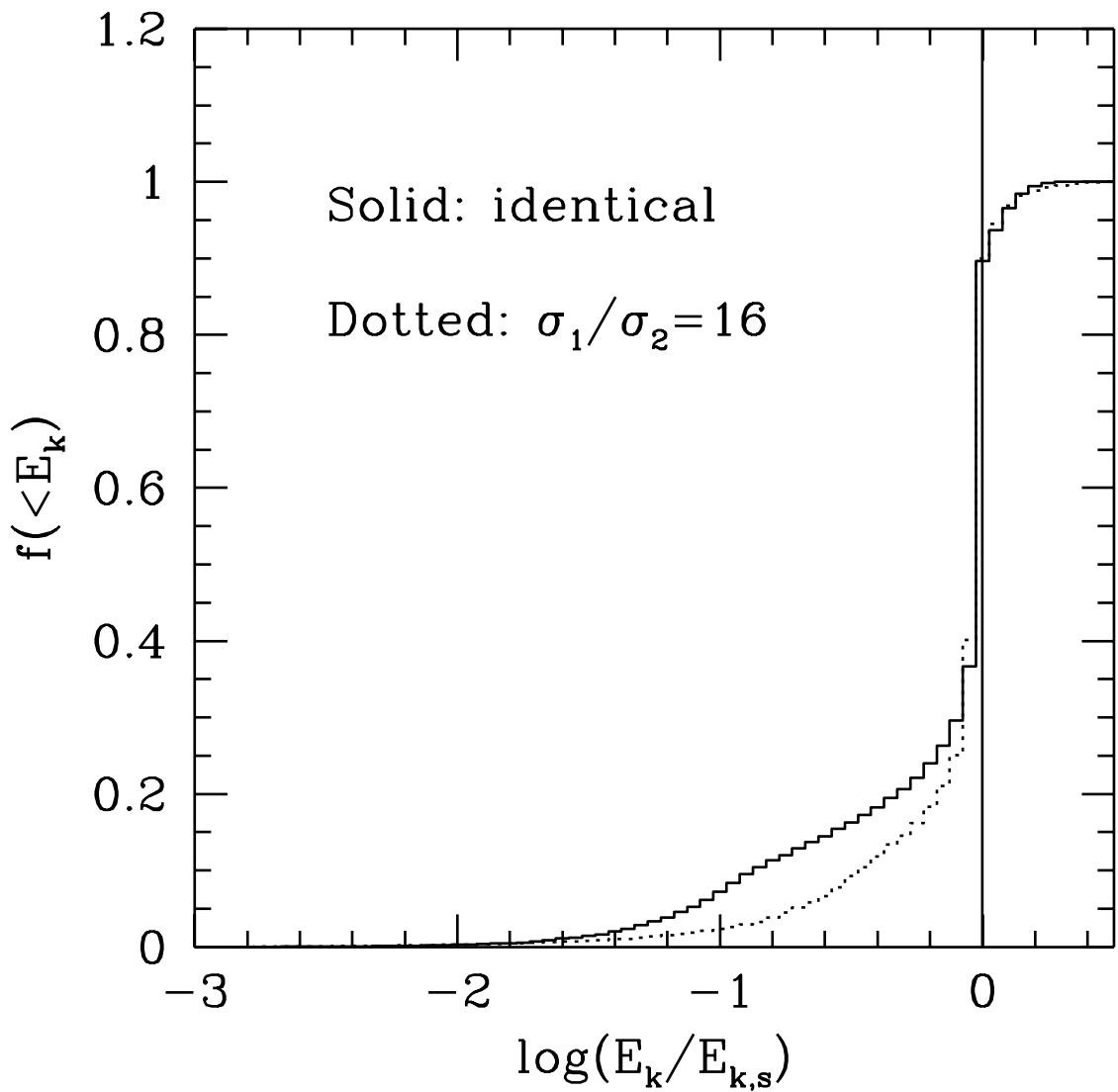